\documentclass[12pt]{iopart}
\usepackage{amsfonts}
\usepackage{amssymb}
\usepackage{latexsym}
\usepackage{graphicx}
\usepackage{dcolumn}
\usepackage{bm}
\usepackage{textcomp}
\usepackage{wasysym}
\usepackage{stmaryrd}
\usepackage{subfig}

\def \ee{\end{equation}}
\def \be{\begin{equation}}
\def \bea{\begin{eqnarray}}
\def \eea{\end{eqnarray}}

\newcommand{\eqref}[1]{(\ref{#1})}

\newcommand{\cR}{\mathcal{R}}

\newcommand{\p}{\partial}

\begin{document}

\title{Black holes and running couplings: \\ A comparison of two complementary approaches
}

\author{Benjamin Koch*, Carlos Contreras$^+$, Paola Rioseco*, and Frank Saueressig**}
 \address{
*Instituto de F\'{i}sica, Pontificia Universidad Cat\'{o}lica de Chile, \\
Av. Vicu\~{n}a Mackenna 4860, Santiago, Chile \\

$^+$Departamento de F\'{i}sica, Universidad T\'{e}cnica Federico Santa Mar\'{i}a;\\
Casilla 110-V, Valpara\'{i}so, Chile\\

**Radboud University Nijmegen, \\ Institute for Mathematics, Astrophysics and Particle Physics (IMAPP),\\
Heyendaalseweg 135, 6525 AJ Nijmegen, The Netherlands \\
}
\date{\today}

\begin{abstract}
Black holes appear as vacuum solutions of classical general relativity which depend on Newton's constant and possibly the cosmological constant. At the level of a quantum field theory, these coupling constants typically acquire a scale-dependence. This proceedings briefly summarizes two complementary ways
to incorporate this effect: the renormalization group improvement of the classical black hole solution based on the running couplings obtained within the gravitational Asymptotic Safety program and the exact solution of the improved equations of motion including an arbitrary scale dependence of the gravitational couplings. Remarkably the picture of the ``quantum'' black holes obtained from these very different improvement strategies is surprisingly similar.
\end{abstract}


\maketitle



%
\section{Introduction}
\label{sec:1}
The emergence of scale-dependent couplings is one of the central phenomena encountered in quantum field theory. While the quest for a consistent and predictive quantum formulation for gravity is still ongoing, it is natural to expect that this feature will emerge in this case as well. 
 This expectation is supported by perturbative computations in the framework of higher-derivative gravity \cite{Stelle:1976gc,Julve:1978xn,Codello:2006in} as well as the non-perturbative computations carried out within the gravitational Asymptotic Safety program \cite{Niedermaier:2006wt,Percacci:2007sz,Codello:2007bd,Reuter:2012id}.

An important testing ground for ideas related to modified theories of gravity or quantum gravity is given by the black hole solutions obtained from classical general relativity. Striving for a quantum description of these objects, it is natural to study the effect of scale-dependent coupling constants on the physics of the black holes. In this proceedings paper we will focus on two complementary strategies for capturing these effects:
\begin{itemize}
\item
The first approach discussed in section \ref{sec:2} was pioneered in \cite{Bonanno:1998ye,Bonanno:2000ep} and performs a renormalization group (RG) improvement
of the classical black hole solution. Here the classical coupling constants are promoted to scale-dependent couplings
whose flow is governed by beta functions computed within Asymptotic Safety. By now, these techniques have been refined by several groups \cite{Emoto:2005te,Bonanno:2006eu,Koch:2007yt,Burschil:2009va,Reuter:2010xb,Falls:2010he,Casadio:2010fw}.
\item 
The second approach covered in section \ref{sec:3} follows the spirit of \cite{Reuter:2003ca} and looks for
consistent solutions of the improved equations of motion.
These equations can be solved without making further assumptions on the actual scale dependence
of the couplings, leading to a new, spherically symmetric metric. This metric can be seen as a promising
candidate for a physical black hole metric that incorporates general effects of scale dependent couplings.
\end{itemize}
In section \ref{sec:4} we will compare those results and conclude.

\section{Improved solutions from Asymptotic Safety}
\label{sec:2}
This section basically follows Ref.\ \cite{Koch:2013owa}. Thus, we restrict ourselves to a summary of the key 
 concepts and results and refer to \cite{Koch:2013owa} for more details and further
references. 

The key ingredient for investigating Weinberg's Asymptotic Safety conjecture \cite{Weinberg:1979} and its phenomenological implications
is the gravitational effective average action $\Gamma_k$ \cite{mr}, a Wilson-type effective action 
that provides an effective description of physics at the momentum scale $k$. As 
its main virtue, the scale-dependence of $\Gamma_k$ is governed by an 
 exact functional renormalization group equation \cite{mr}
\be\label{FRGE}
\p_k \Gamma_k = \frac{1}{2}  {\rm Tr}  \left[ \left( \Gamma_k^{(2)} + \cR_k \right)^{-1} \, \p_k \cR_k \right] \, .
\ee
Here $\Gamma_k^{(2)}$ denotes the second variation of $\Gamma_k$ with respect to the quantum fields and $\cR_k$ is an IR-regulator that renders the trace finite and peaked on fluctuations with momenta $p^2 \approx k^2$.

\begin{figure}[t]
\centering
\includegraphics[width=0.6\textwidth]{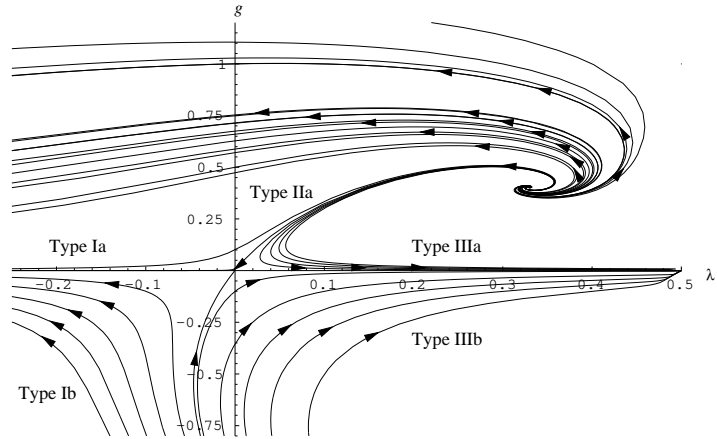}
\caption{RG flow originating from the Einstein-Hilbert truncation \eqref{action}. The arrows point in the direction of increasing 
coarse-graining, i.e.\ of decreasing $k$. From \cite{Reuter:2001ag}. \label{EHflow}}
\end{figure}

The simplest setup for obtaining a non-perturbative approximate solution of \eqref{FRGE} truncates
the gravitational part of $\Gamma_k$ to the (scale-dependent) Einstein-Hilbert action
 \be\label{action}
 \Gamma_k^{\rm grav}[g] = \frac{1}{16 \pi G_k} \int d^4x \sqrt{g} \left[-R + 2 \Lambda_k \right] \, , 
 \ee
 which includes two running couplings, Newton's constant $G_k$ and the cosmological constant $\Lambda_k$.
 The beta functions resulting from this truncation have first been derived in \cite{mr} and
are most conveniently expressed in terms of the dimensionless 
 coupling constants
\be\label{Gvong}
 g_k= G_k \, k^{2}\quad, \quad
\lambda_k= \Lambda_k \, k^{-2} \quad \, . 
\ee
The phase diagram resulting from the flow  has been constructed in \cite{Reuter:2001ag} and is shown 
in figure \ref{EHflow}. 
The flow is governed by the interplay of a Gaussian fixed point located at the origin, $g_* = 0, \lambda_* = 0$
and a non-Gaussian fixed point (NGFP) governing the UV-behavior of the flow. For the optimized cutoff this NGFP is located
at 
\be\label{NGFP}
\lambda_*=0.193 \, , \qquad g_*=0.707 \, , \qquad g_* \lambda_* = 0.137\quad.
\ee

One way to investigate the implications of the scaling gravitational couplings on (A)dS black holes 
is the RG improvement of the classical black hole solution. 
This procedure starts from the classical (Schwarzschild-de Sitter or anti-de Sitter) 
line-element
\be\label{lineele}
ds^2= -f(r) \, dt^2+ f(r)^{-1} \, dr^2 + r^2 d\Omega_2^2
\ee
with
\be\label{frfct}
f(r) = 1 - \frac{2 G M}{r} - \frac{1}{3} \, \Lambda \, r^2 \, ,
\ee
and replaces Newton's constant and the cosmological constant by their scale dependent counterparts, 
$G\rightarrow G_k$, $\Lambda \rightarrow \Lambda_k$.
The crucial step following this improvement is the scale setting procedure,
which relates the momentum scale $k$ to the radial scale $r$
\be\label{kvonP}
k(P(r))=\frac{\xi}{d(P(r))}\quad,
\ee
where $\xi$ is an a priory undetermined constant. On general grounds
the cutoff identification $d(P)$ should be independent of
the choice of coordinates and compatible with the
symmetries of the classical solution. 
Following \cite{Bonanno:2000ep}, a natural candidate for $d(P)$
is the radial proper distance between the point $P$ and the origin which 
should provide the physical cutoff of the geometry.

Applying this improvement scheme to the classical 
(A)dS black holes led to various novel conclusions, which are largely independent
of the details underlying the scale setting procedure:
\begin{itemize}
\item[a)] Including the effect of a scale-dependent cosmological constant in the RG-improvement process drastically affects the structure of the quantum-improved black holes \emph{at short distances}.
Thus a consistent RG-improvement procedure requires working in the class of Schwarzschild-(A)dS solutions of Einstein's equations.
\item[b)]
The short-distance structure of all quantum-improved black holes is governed by the NGFP. This entails that the structure of light black holes is universal. In particular it is independent of the IR-value of Newton's constant and the cosmological constant and therefore identical for classical Schwarzschild, Schwarzschild-dS and Schwarzschild-AdS black holes. 
\item[c)]
In the presence of the cosmological constant, the curvature singularity at $r = 0$ is not resolved.
\end{itemize}
%

\section{Solving improved equations of motion}
\label{sec:3}
An alternative strategy for modeling the quantum properties 
of a classical black hole, based on ``improving the equations of motion'',
has been developed in \cite{Contreras:2013hua}. In this case, the scale-setting
procedure is carried out at the level of the (wick-rotated) Einstein-Hilbert
action \eqref{action} where the $k$-dependence of the couplings is replaced by
a generic $r$-dependence. The resulting equations of motion
are \cite{Reuter:2004nx,Domazet:2012tw}
\begin{eqnarray}
 G_{\mu\nu}=-g_{\mu\nu}\Lambda(r)+ 8 \pi G(r) T_{\mu \nu}
 -\Delta t_{\mu \nu}\quad,
\label{eom2}
\end{eqnarray}
with
\be
\Delta t_{\mu \nu}=G(r)\left(g_{\mu\nu}\Box-\nabla_\mu\nabla_\nu\right)\frac{1}{G(r)}\quad.
\ee
With the metric ansatz
\be\label{ansatz}
ds^2=-F(r) dt^2+
1/F(r)dr^2+
r^2 d\theta^2
+r^2 \sin (\theta)d\phi^2\quad,
\ee
the equations of motion can be solved exactly, 
for the functions $F(r), \; \Lambda(r)$, and $G(r)$. This solution is non-trivial, leading to four constants of integration 
$c_1,\;c_2,\;c_3,\;c_4$. These constants can be related
to familiar properties of the classical solution such as $M_0$, $G_0$, $\Lambda_0$ together with a possible correction. 
Alternatively, they can be traded for the adimensional parameters $g_I$, $g_U$, $\lambda_I$, and $\lambda_U$ which naturally appear
in the induced coupling flow
%
\bea\label{UVFP}
g_{U}(r)&=& G(r) \Sigma^2 \, , \qquad
\lambda_{U}(r) = -\Lambda(r) \frac{r}{\Sigma} \quad,
\eea
where $\Sigma$ is an arbitrary matching constant which has mass dimension one.
\begin{figure}[t]
\centering
\includegraphics[width=0.6\textwidth]{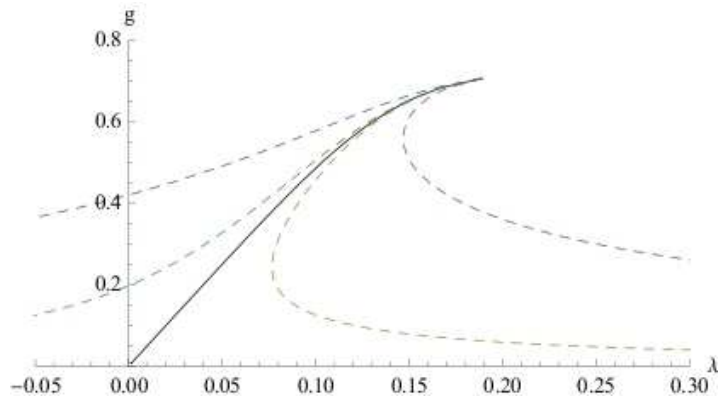}
\caption{Schematic flow of the scale dependent couplings $\lambda_{U}(r)$ and $g_{U}(r)$ for $g^*_U=0.707$,
  $\lambda^*_U=0.193$, $g_I=2.5$, and $G_0=\Sigma=1$. The different curves correspond to
  $l_I=\{ -0.05,\,-0.005,\,0,\,0.005,\,0.05\}$. From \cite{Contreras:2013hua}. \label{indflow}}
\end{figure}
The values of the UV fixed points of this ``flow''  are
\bea\label{UVFPlim}
g_{U}(r\rightarrow 0)&=&g_{U}^* \, , \qquad
\lambda_{U}(r\rightarrow 0) = \lambda_{U}^* \quad.
\eea
The induced ``flow'' for the couplings \eqref{UVFP} is shown in figure~\ref{indflow} and
turns out to be surprisingly similar to the genuine RG flow shown in figure~\ref{EHflow}.
Moreover, the main properties of the improved solutions can be summarized as follows
\begin{itemize}
\item[a)] There exists a non-trivial solution of the improved equations of motion (\ref{eom2})
which can not be obtained without the cosmological term. Thus, including a scale
dependent cosmological term is crucial for this approach.
\item[b)] In the UV limit, the new solution is dominated by the fixed points $g_U^*$ and $\lambda_U^*$.
\item[c)] The new solution $F(r)$ exhibits a singularity at the origin, which is of the same grade
as the singularity of the Schwarzschild solution. In this limit the
solution is dominated by the non-trivial fixed point of the induced ``flow''.
\item[d)] Interpreting the solution for the couplings (\ref{UVFP}) as  flow 
one finds interesting similarities with the RG flow derived from the effective average action $\Gamma_k$.
\end{itemize}

\section{Comparison of the two approaches}
\label{sec:4}
We have summarized two strategies for modeling quantum
corrections to classical black hole solutions based on
implementing scale-dependent couplings.
The first approach is based on improving the classical solutions 
and uses the beta functions obtained from Asymptotic Safety
to fix the scale-dependence of the gravitational coupling
constants (see section \ref{sec:2}). The second approach is based on simply solving
the equations of motion that have to be fulfilled in the presence of scale dependent couplings $\Lambda(r)$ and $G(r)$
in a generally covariant theory (\ref{eom2}).
These a priori unrelated schemes lead to a qualitatively similar picture
for the improved black hole solutions:
\begin{itemize}
\item[a)] {\bf $\Lambda$ matters:}\\
In both approaches the cosmological constant has a significant effect. 
In section \ref{sec:2} this term strongly dominated the UV behavior of the improved solution, while
in section \ref{sec:3} this term was actually crucial for obtaining a non-trivial solution at all.
\item[b)] {\bf Fixed points control UV:}\\
In both approaches the short distance behavior is dominated by the non-trivial fixed point
of the true flow in section \ref{sec:2} or of the induced ``flow'' in section \ref{sec:3}. 
\item[c)]{\bf Singularity persists:}\\
Rather surprisingly, both approaches exhibit the same type of black hole singularity 
located at the origin. Since it is a general expectation
that quantum gravity should be capable of resolving
this singularity, it would be very interesting
to understand this result in more detail.
\end{itemize}
These coincidences consolidate the findings of both approaches.

\section*{Acknowledgements}
B.K.\ thanks the organizers of the first Karl Schwarzschild meeting for hospitality. The work of B.K.\ was supported proj.\ Fondecyt 1120360 
and anillo Atlas Andino 10201; the research of F.S.\ is
supported by the Deutsche Forschungsgemeinschaft (DFG)
within the Emmy-Noether program (Grant SA/1975 1-1).
The work of C.C.\ was supported proj.\ Fondecyt 1120360 and DGIP grant  11.11.05.

\bigskip

\end{document}